# Specific heat of unconventional superconductivity in $Na_xCoO_2 \cdot yH_2O$

H. D. Yang[1], J.-Y. Lin[2], C. P. Sun[1], Y. C. Kang[1], and K. Takada[3], T. Sasaki[3], H. Sakurai[4] and E. Takayama-Muromachi[4]

[1]*Department of Physics, National Sun Yat Sen University, Kaohsiung 804, Taiwan, R.O.C.*

[2]*Institute of Physics, National Chiao Tung University, Hsinchu 300, Taiwan, R.O.C.*

[3]*Soft Chemistry Research Group, Advanced Materials Laboratory, National Institute for Materials Science, 1-1 Namiki, Tsukuba, Ibaraki 305-0044, JAPAN*

[4]*Superconducting Materials Center, National Institute for Materials Science 1-1 Namiki, Tsukuba, Ibaraki 305-0044, Japan*

Comprehensive studies of the low-temperature specific heat $C(T,H)$ in $Na_xCoO_2 \cdot yH_2O$ are presented. At $H=0$, a very sharp anomaly was observed at $T=4.7$ K indicating the existence of bulk superconductivity. There exists the $rT^2$ term in $C(T,H=0)$ in the superconducting state manifesting the line nodal superconducting order parameter. The superconducting volume fraction is estimated to be 26.6 % based on the consideration of entropy conservation at $T_c$ for the second-order superconducting phase transition. An abrupt change of the slope in $T_c(H)$ was observed. Possible scenarios such as the multiple phase transitions in the mixed state are discussed.

The newly reported superconductivity in $Na_xCoO_2 \cdot yH_2O$ with $T_c \leq 5$ K [1,2] generates new excitement in condensed matter physics community. Though its $T_c$ is much lower than that of cuprate superconductors, $Na_xCoO_2 \cdot yH_2O$ has stimulated broad interest for several reasons. The parent compound $Na_xCoO_2$ has been known to be a strongly correlated system [3,4] with the triangular $CoO_2$ two-dimensional (2D) sublattice. The system could be considered as an electron-doped Mott insulator through sodium doping. Similar to the importance of the strong 2D character in high-$T_c$ cuprates, the large separation of the $CoO_2$ layers by the intercalation of $H_2O$ molecules seems to be essential for inducing the superconductivity in $Na_xCoO_2 \cdot yH_2O$. This model is consistent with the recent negative hydrostatic pressure effects on the $T_c$ of the present material [5]. Consequently, the theoretical interest is straightforward and obvious. Elucidation of the superconducting mechanism in cuprate superconductors could be improved by studying this similar system. On the other hand, with the triangular $CoO_2$ planes rather than the nearly square $CuO_2$ planes, there possibly exists new superconductivity and an alternative to reach high-$T_c$. Theoretical models with unconventional superconductivity have been proposed [6-8]. However, the fundamentals of $Na_xCoO_2 \cdot yH_2O$ are far from being experimentally established at this moment. For example, several reports of NMR and NQR experiments have reached different conclusions on the order parameter symmetry [9-12]. The specific heat ($C$) technique can probe the bulk properties of the samples and has been proven to be a powerful tool to investigate the pairing state of novel superconductors such as high-$T_c$ cuprates [13-16], $MgB_2$ [17,18], and $MgCNi_3$ [19]. $C(T,H)$ also provides the information about the quasiparticle excitation associated with the vortex state. In this paper, we present comprehensive low-temperature specific heat studies of $Na_xCoO_2 \cdot yH_2O$. The results imply an unconventional pairing symmetry in $Na_xCoO_2 \cdot yH_2O$. In magnetic fields $H$, an anomalous $T_c(H)$ curve is observed suggesting the complicated magnetic correlation in the superconducting state. Polycrystalline $Na_xCoO_2 \cdot yH_2O$ powder was prepared and characterized as described in [1]. The composition was determined to be $x=0.35$ and $y=1.3$. Thermodynamic $T_c$ determined from $C(T)$ is 4.7 K (see below) consistent with that observed from the magnetization measurements [1]. $C(T)$ was measured using a $^3He$ thermal relaxation calorimeter from 0.6 K to 10 K in magnetic fields $H$ up to 8 T. A detailed description of the measurements can be found in Ref. [18]. The powder was cold pressed into pellets applying a pressure of about $1.6 \times 10^4$ nt $cm^{-2}$ for $C$ measurements. The samples had been exposed to air with humidity above 50% for about 15 minutes or less during the procedure before being cooled down to low temperatures in a helium gas environment. Since the pellets were cold pressed rather than hot pressed, extra caution was taken on the thermal conductivity of the sample. Only pellets thinner than 0.5 mm were used in $C$ measurements. The background contribution (from the addendum plus grease) was separately measured and subtracted from the data. One of the samples was measured two days after the first run of the specific heat measurements. Both runs rendered identical $C(T)$ within the apparatus resolution limit, indicating the stability of the samples in liquid helium temperature.

The data of $C(T,H=0)$ for one sample are shown as $C/T$ vs. $T^2$ in Fig. 1. A pronounced peak of $C(T)/T$ manifests the bulk phase transition occurring at $T_c=4.7$ K in $Na_xCoO_2 \cdot yH_2O$. Indeed, $C/T$ does not extrapolate to zero as $T$ approaches zero, which suggests that only a portion of the sample undergoes the superconducting phase transition. However, the peak is as sharp as those observed in many other well identified superconductors [17-20]. Therefore, the existence of a well separated superconducting portion in the sample rather than a broad spread in $T_c$ can be taken as a plausible assumption. If the peak in $C/T$ is due to the superconducting transition, integration of $\delta C_e/T \equiv C(H=0)/T - C_n/T$ from $T=0$ to $T_c$ should be zero owing to the requirement of entropy conservation. Here $C_n$ is the normal state specific heat and can be written as $C_n(T)=\chi_n T+C_{lattice}$, where $C_{lattice}=ST^3+DT^5$ represents the phonon contribution. Naively, one may try to obtain $C_n$ by fitting the data above $T_c$. A more elaborate analysis is to take the conservation of entropy into consideration and thus naturally includes the low temperature $C_n$ below $T_c$, which was not directly measured due to large $H_{c2}$ in this sample. This further analysis results in $\chi_n=14.89$ mJ/mol $K^2$, $S=0.110$ mJ/mol $K^4$ (the Debye temperature $\Theta_D=503$ K) and $D=6.89 \times 10^{-4}$ mJ/mol $K^6$, shown as the solid curve in Fig. 1. The entropy balance is achieved as shown in the inset of Fig. 1. It is noted that failing to include the $uT^5$ term would lead to severe entropy imbalance as in some of other related works [21-25].

Fruitful information of the superconductivity in $Na_xCoO_2 \cdot yH_2O$ can be deduced from $\delta C_e(T)/T$ shown in Fig. 2. First, $\delta C_e(T)/T$ below $T=2$ K is linear with respect to $T$. This behavior strongly suggests an $rT^2$ term ($r$ is a constant) in the electronic specific heat $C_e$ below $T_c$. At higher temperatures, $\delta C_e(T)/T$ gradually transits from $rT^2$ to a faster increase, and apparently affected by the superconducting anomaly near $T_c$. This $rT^2$ term is a manifestation of the nodal lines in the superconducting order parameter. To further check the existence of an $rT^2$ term, $C(T)$ of another sample was measured down to 0.6 K at $H=0$. As seen in the inset of Fig. 2, the nearly identical $\delta C_e(T)/T$ deduced from two different samples warrants the measurements and analysis in the present paper. The overall data of $\delta C_e(T)/T$ in the inset of Fig. 2 *follow the linear T dependence down to the lowest measurement temperature*. Consequently, the $rT^2$ term seems to be robust and is sample independent. This observation is in sharp contrast with the $T^3$ dependence reported in Ref. [21]. Linear fit between 1K and 2 K (the dashed line in Fig. 2) leads to $\delta C_e(T=0)/T$=-3.96 mJ/mol K$^2$. The specific-heat jump $(\Delta C/T_c)_{ob}$=7.79 mJ/mol K$^2$ at $T_c$ is determined by the entropy balance near $T_c$ as shown in Fig. 2. However, the corresponding value of $\chi_n$ should be appropriately taken as 3.96 mJ/mol K$^2$, which is from the carriers participating the superconducting transition, rather than 14.89 mJ/mol K$^2$ which includes additional contribution from the nonsuperconducting part. Therefore, the *normalized* dimensionless specific-heat jump $\Delta C/\chi_n T_c$=1.96. In BCS weak–coupling limit, $\Delta C/\chi_n T_c$=1.43 and ~1 respectively for the superconductivity of isotropic $s$-wave and of the order parameter with line nodes [26]. Therefore, the observation of the $rT^2$ term together with the value of $\Delta C/\chi_n T_c$ implies that $Na_xCoO_2 \cdot yH_2O$ is likely a strong-coupling superconductor with nodal lines in the order parameter. This strong coupling is also consistent with the large mass enhancement resulting from the comparison between the observed $\chi_n$ and the density of states of the band calculations [23,27]. By the similar analysis as above, the volume fraction of the superconducting portion is estimated by $(-\delta C_e(T=0)/T)/\chi_n$=3.96/14.89=26.6%. This number is consistent with that from the magnetization measurements, by which up to 20% of superconducting fraction was estimated [1,28]. Moreover, it is of interest to briefly discuss the value of $r$. The estimated coefficient $r \approx \chi_n/T_c$, with a prefactor of order 1 depending on the details of the Fermi surface. For $Na_xCoO_2 \cdot yH_2O$, the observed $r$=1.02 mJ/mol K$^3$ (from the slop of the dashed line in Fig. 2) is in good accord with the estimated $r \approx 3.96/4.7$=0.85 mJ/mol K$^3$. Similar agreement was also observed in other line nodal superconductors $La_{1.78}Sr_{0.22}CuO_4$ and $Sr_2RuO_4$ [14,26]. The $rT^2$ term actually appears in all the superconducting $Na_xCoO_2 \cdot yH_2O$ samples we have measured. Even if $C_n(T)$ is naively determined from Fig. 1 between 7 to 10 K and a fit of $C_n(T)$ with only the $sT^3$ term in $C_{lattice}$ is implanted, The $rT^2$ term is still robust regardless of the entropy imbalance in this kind of analysis. On the other hand, since the nonsuperconducting volume in the sample is larger than the superconducting one, there might be sources of uncertainty in the $rT^2$ term. Furthermore, if there possibly exists a small portion of the sample with much lower $T_c \ll 4.7$, the above interpretation of the $rT^2$ term could be complicated.

$C(T,H)/T$ in magnetic fields can be seen in Fig. 3. Although $H$ seems to strongly suppress the anomaly peak, $T_c(H)$ actually decreases with $H$ rather slowly as shown in Fig. 4. In the inset of Fig. 4, taking the data in $H=1$ T as an example, two straight lines are extrapolated from the measured $C(T)/T$ data with $T$ just above or below the transition, and $T_c$ is determined by the local entropy balance around the transition. The error bar is determined by the entropy imbalance of about 50% of the entropy enclosed by the data, the straight line at $T_c$ and one of the extrapolated lines. This intriguing $H$ dependence of $T_c$ in Fig. 4 is amazingly consistent with the results of the magnetization measurements independently performed by different group on polycrystalline samples from different sources [1,28,29]. Most noticeable, there is a change of the slope in the $T_c$-$H$ curve at $H\sim 1$ T, suggesting that an $H$-induced phase transition, probably from the nodal to $s$-wave symmetry or from the singlet to triplet pairing, occurs at high $H$. The later possible transition was theoretically interpreted in Ref. [29] by the resistivity measurements. Another possible source of the slope change could be the anisotropy of $H_{c2}$ as reported in Ref. [23]. Furthermore, this slow decrease in $T_c(H)$ for $H \gtrsim 1$ T implies an $H_{c2}$ probably higher than 50 T. For a superconductor with $T_c=4.7$ K, such a high $H_{c2}$ is certainly unusual. Another intriguing result is that the magnetic field has nearly no suppression effect on the onset temperature $T_{on}$ of the transition for $H<4$ T as shown in the inset of Fig. 3. This might imply that fluctuations have strong influence on $T_c$ and/or $T_{on}$ and could complicate the determination and discussions on $T_c(H)$ in Fig. 3.

At low temperatures, it is noted that $C(T,H)/T$ first increases with increasing $H$ and then decreases for $H>4$ T. This nonmonotonic behavior is likely due to the $C_{\text{Schottky}}(g\sim H/k_BT)$ contribution from the Schottky anomaly due to the paramagnetic centers in samples. Furthermore, at least some of the contribution in the broad anomaly in $C/T$ at $H=8$ T is from the Schottky anomaly. This Schottky term is also partially responsible for the extra $C/T$ contribution in $H$ above $T_c=4.7$ K as shown in Fig. 3. If the superconductivity is a second order phase transition, thermodynamics requires

$$\int_0^{T_c} \frac{uC_e(H,T)}{T} dT = 0$$

(1)

as approximately observed in many superconductors [17-20]. However, this conservation law is violated in the present sample, at least partially due to the presence of the paramagnetic centers. A rough estimate of the low temperature $C(T,H)$ data leads to a paramagnetic center concentration higher than $10^{-3}$, which is not too surprising in a cobalt oxide. The strong Schottky anomaly virtually hinders the reliable investigation of $\chi(H)$, which is valuable to the understanding of the superconductivity. Therefore, further studies of C(T, H) in cleaner $Na_xCoO_2 \cdot yH_2O$ samples free from paramagnetic centers are desirable.

In Summary, Comprehensive studies of the low-temperature specific heat $C(T,H)$ in $Na_xCoO_2 \cdot yH_2O$ are presented. At $H=0$, a very sharp anomaly was observed at $T=4.7$ K indicating the existence of bulk superconductivity. There exists the $T^2$ term in $C(T,H=0)$ in the superconducting state manifesting the line nodal superconducting order parameter. The dimensionless specific-heat jump $\Delta C/\gamma_n T_c$, probable as large as 1.96, indicates strong coupling in superconducting $Na_xCoO_2 \cdot yH_2O$. The results of $T_c(H)$ suggest unconventional magnetic properties, and multi superconducting phases in $H$ are implied.

We are grateful to C. Y. Mou and T. K. Lee for indispensable discussions. Experimental help from L. S. Chao is appreciated. This work was supported by National Science Council, Taiwan, Republic of China under contract Nos. NSC92-2112-M-110-017 and NSC92-2112-M-009-032.

**Figure captions**

Fig. 1.  $C/T$ vs. $T^2$ for $Na_xCoO_2 \cdot yH_2O$. The solid curve represents the normal state $C_n/T = \chi_n + ST^2 + uT^4$. Inset: entropy difference $\Delta S$ by integration of $\delta C_e(T)/T$ according to the data above 1 K and the dashed line shown in Fig. 2 below 1 K.

Fig.2. $\delta C_e/T \equiv C(H=0)/T - C_n/T$ vs. $T$ for $Na_xCoO_2 \cdot yH_2O$. The dashed line shown in both the main figure and in the inset represents the linear fit of the solid circle data between 1 and 2 K. Inset: $\delta C_e/T$ of two samples at low temperatures clearly showing the linear $T$ dependence. Open circle data measured down to 0.6 K are from another sample.

Fig. 3.  $C/T$ vs. $T^2$ for $Na_xCoO_2 \cdot yH_2O$ in magnetic fields $H$. For clarity, only data at selected fields are shown. Inset: $C/T$ vs. $T$ for $H \leq$ 1T showing that $T_{on}$ nearly does not change in $H$.

Fig. 4.  $T_c(H)$ thermodynamically determined from $C(T,H)$. The dashed lines are guides to eyes. Inset shows the example for $H=1$ T data how $T_c(H)$ is determined by the local entropy balance.

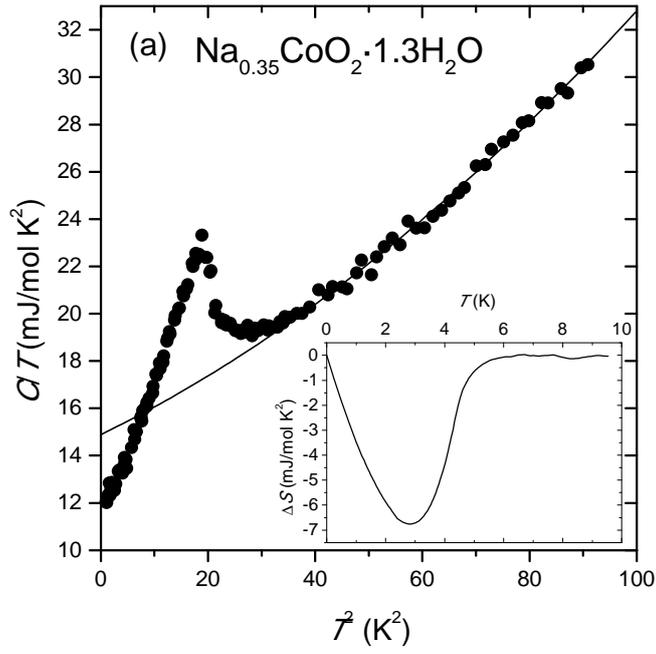

Fig. 1 Yang et al.

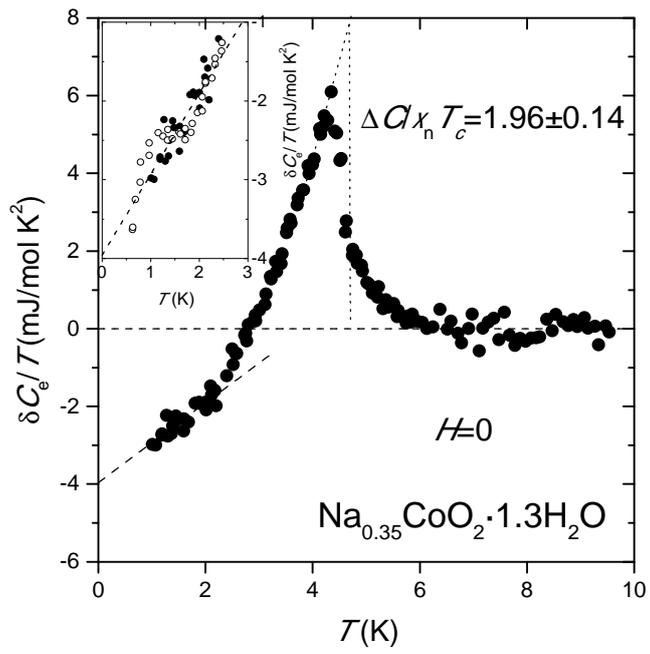

Fig. 2 Yang et al.

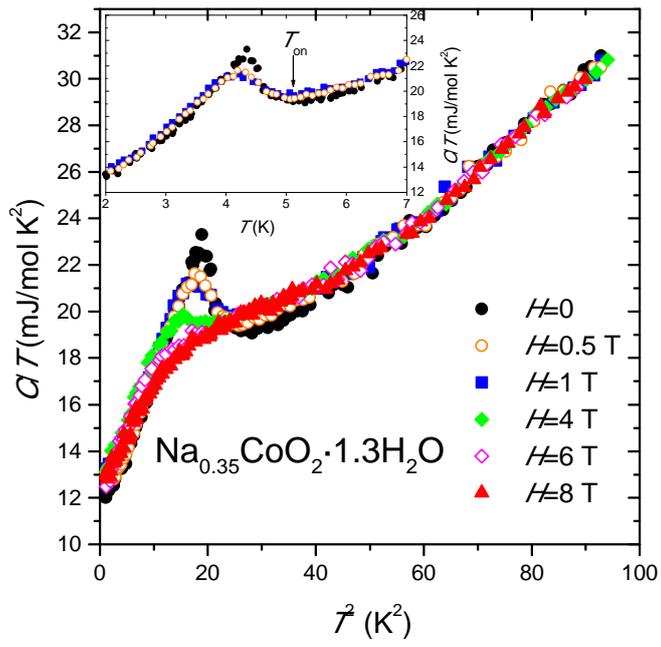

Fig. 3 Yang et al.

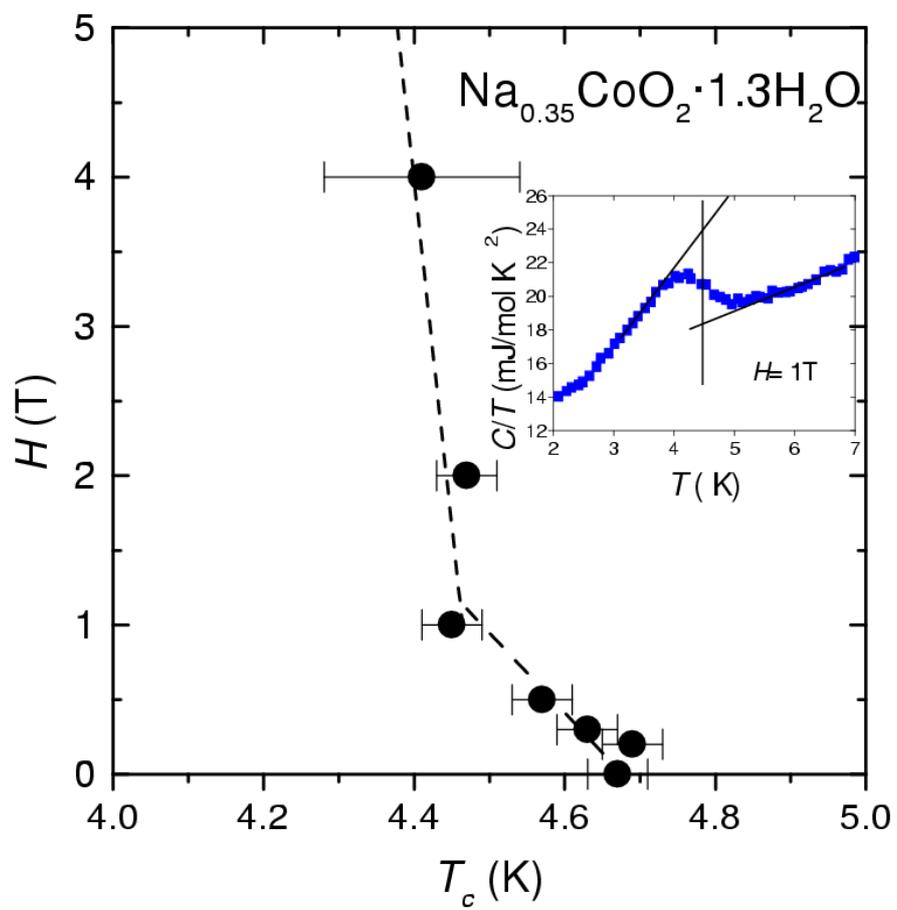

Fig. 4 Yang et al.